\documentclass[lettersize,journal]{IEEEtran}
\usepackage{algorithmic}
\usepackage{algorithm}
\usepackage{array}
\usepackage[caption=false,font=normalsize,labelfont=sf,textfont=sf]{subfig}
\usepackage{textcomp}
\usepackage{stfloats}
\usepackage{url}
\usepackage{verbatim}
\usepackage{graphicx}
\usepackage{cite}
\usepackage{amsmath,amssymb,amsfonts}
\usepackage{xcolor}
\usepackage{hyperref}

\begin{document}

% \title{A Haptic Device Recommendation System Leveraging Context-Aware LLM Agents}
\title{Leveraging LLMs to Create a Haptic Devices' Recommendation System}
% Leveraging LLMs Agent for Context-Aware Haptic Device Recommendation

\author{Yang Liu,~\IEEEmembership{Member,~IEEE,} Haiwei Dong,~\IEEEmembership{Senior Member,~IEEE,} and Abdulmotaleb El Saddik,~\IEEEmembership{Fellow,~IEEE}
\thanks{Y. Liu, H. Dong, and A. El Saddik are with the Multimedia Computing Research Laboratory (MCRLab), School of Electrical Engineering and Computer Science, University of Ottawa, Ottawa, ON K1N 6N5, Canada. E-mail: \{yliu344; hdong;  elsaddik\}@uottawa.ca}
}

\markboth{Journal of \LaTeX\ Class Files,~Vol.~14, No.~8, August~2021}%
{Shell \MakeLowercase{\textit{et al.}}: A Sample Article Using IEEEtran.cls for IEEE Journals}


% \author{Yang Liu\\
% University of Ottawa\\
% IEEE Member\\
% {\tt\small yliu344@uottawa.ca}
% % Additional authors and addresses can bare added with ``\and'',
% \and
% Haiwei Dong\\
% Huawei Canada, University of Ottawa\\
% IEEE Senior Member\\
% % First line of institution2 address\\
% {\tt\small hdong@uottawa.ca}
% \and
% Abdulmotaleb El Saddik\\
% University of Ottawa, MBZUAI\\
% IEEE Fellow\\
% {\tt\small elsaddik@uottawa.ca}
% }

\maketitle

\begin{abstract}
Haptic technology has seen significant growth, yet a lack of awareness of existing haptic device design knowledge hinders development. This paper addresses these limitations by leveraging advancements in Large Language Models (LLMs) to develop a haptic agent, focusing specifically on Grounded Force Feedback (GFF) devices recommendation. Our approach involves automating the creation of a structured haptic device database using information from research papers and product specifications. This database enables the recommendation of relevant GFF devices based on user queries. To ensure precise and contextually relevant recommendations, the system employs a dynamic retrieval method that combines both conditional and semantic searches. 
Benchmarking against the established UEQ and existing haptic device searching tools, the proposed haptic recommendation agent ranks in the top 10\% across all UEQ categories with mean differences favoring the agent in nearly all subscales, and maintains no significant performance bias across different user groups, showcasing superior usability and user satisfaction.
% Experimental results using the UEQ framework indicate that our haptic agent outperforms existing systems in key dimensions of usability and satisfaction, with participants rating the system highly across all subscales.
% comprises two primary subsystems: the Creator System, which automates the creation of a structured haptic device database from research papers and product specifications, and the User System, which interprets user queries and recommends relevant GFF devices. The User System employs a dynamic Retrieval-Augmented Generation (RAG) method that integrates both conditional and semantic searches to provide precise and contextually relevant recommendations. 
\end{abstract}

% This paper addresses this challenge by proposing a novel LLM-based system for Grounded Force Feedback (GFF) device selection. The system leverages an LLM's ability to understand user needs and search a central repository of device information. This facilitates efficient GFF device selection for creators, empowering them to develop richer touch experiences. We introduce a novel Retrieval-Augmented Generation (RAG) retrieval method specifically designed for structured data like GFF specifications. Our system operates in Creator and User Modes, allowing for populating the database and retrieving relevant devices based on user descriptions. We demonstrate the system's effectiveness by expanding the database to 114 entries and creating a user-friendly online application.

\begin{IEEEkeywords}
Haptics, Haptic Devices, Large Language Models, Retrieval-Augmented Generation, Grounded Force Feedback
\end{IEEEkeywords}

\begin{figure}
\centering
 \includegraphics[width=.51\textwidth, trim={18 25 15 25},clip]{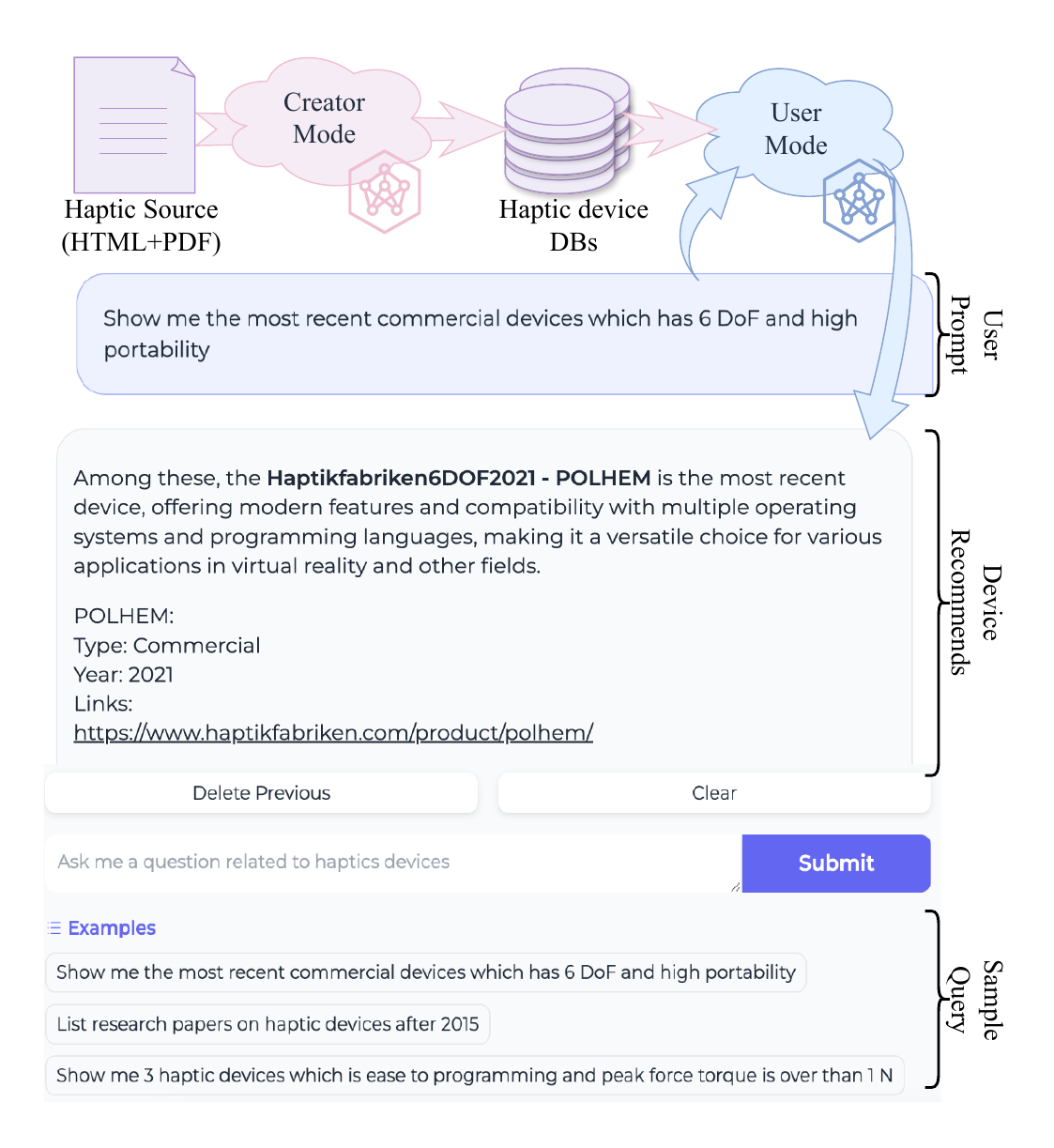}
   \caption{An snapshot for LLM based Haptic device recommendation system, which mainly composed by 3 aspects: 1) User Prompt: The user input prompt to inquiry the potential haptic devices; 2) Device Recommendation: The LLM responds to the user prompt, which ismainly composed by the readable device recommendation and relative detail device info with link. 3) User input box and sample queries.}
\label{fig:high_level}
% \vspace{-0.3cm}
\end{figure}

\section{Introduction}
The field of haptics has experienced significant growth, expanding applications across various domains \cite{adilkhanov2022haptic, dong2015development, Survey2023Tong, Classification2023Li, Robust2024Khojasteh}. However, the dissemination and accessibility of haptic device design knowledge remain limited due to fragmentation across disciplines and complex technical descriptions. This fragmentation poses challenges to the development of effective kinesthetic experiences. Establishing a system to share and reuse existing haptic content could address this issue, thereby benefiting the field as a whole.

Several initiatives, such as Haptipedia \cite{seifi2019haptipedia} and Haptiverse \cite{samithamby2021towards}, have emerged to bridge this gap. These platforms allow hapticians from diverse backgrounds to contribute and access valuable information and resources. Despite their utility, these repositories still face limitations, such as the time-consuming and inefficient process of identifying suitable devices and extracting source haptic device data from original research papers or device specifications.

% The field of haptics has grown significantly with applications in various domains \cite{adilkhanov2022haptic,dong2015development,sigrist2013augmented}. However, despite the increasing interest, there is a lack of awareness of existing haptic device design knowledge due to fragmentation across disciplines and technical descriptions. This challenge hinders the development of effective touch experiences. Sharing and reusing existing haptic content through a central platform could address this issue and benefit the field.

% Several initiatives have emerged to bridge this knowledge gap, such as Haptipedia \cite{seifi2019haptipedia} and the Haptiverse \cite{samithamby2021towards}. These applications allow hapticians from various backgrounds to contribute and access information and resources. However, these platforms still face limitations. Finding the right device and creating source haptic device data from original papers or device information can be a time-consuming and inefficient process.

\begin{figure*}
\centering
\includegraphics[width=.95\textwidth, trim={20 11 22 8},clip]{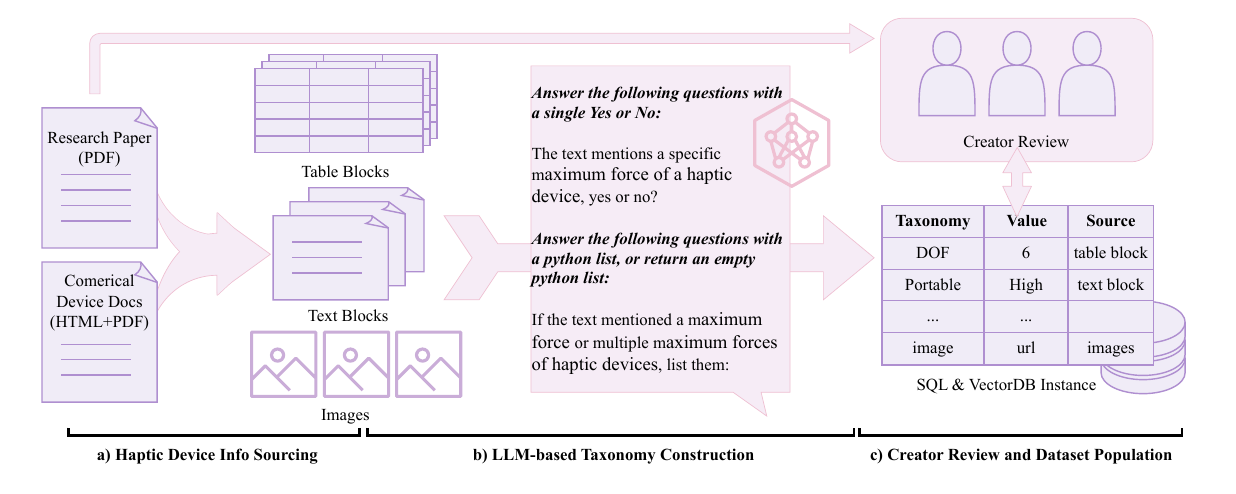}
% \vspace{-0.3cm}
   \caption{The pipeline of the creator-centric system of haptic agent to convert haptic device information from raw source materials into a structured database. Steps a)-d) represent the stages within the pipeline, with detailed descriptions available in Section \ref{sec:creator}1-3) respectively.}
\label{fig:framework}
% \vspace{-0.3cm}
\end{figure*}

One promising solution to this problem is the development of a Large Language Model (LLM)-based haptic hardware recommendation system. LLMs are advanced AI systems trained on vast amounts of text data, enabling them to understand and generate human-like text in response to a wide array of prompts and questions \cite{touvron2023llama, achiam2023gpt}. While LLMs have revolutionized fields such as natural language processing and code generation \cite{touvron2023llama, roziere2023code}, their application in haptics remains largely unexplored. By integrating a central repository of haptic device information with the capabilities of an LLM to understand user needs and search through the data, we can streamline the haptic hardware recommendation process for designers. This integration would bridge the existing knowledge gap and empower creators to develop richer and more effective tactile experiences.

% One promising solution is the development of a Large Language Model (LLM) \cite{touvron2023llama,achiam2023gpt} based haptic hardware selection system. LLMs are powerful AI systems trained on massive amounts of text data, enabling them to communicate and generate human-like text in response to a wide range of prompts and questions \cite{gupta2024rag}. While LLMs have revolutionized fields like natural language processing and code generation \cite{touvron2023llama,roziere2023code}, their application in haptics remains unexplored. A central repository containing information on various haptic devices, combined with an LLM's ability to understand user needs and search through this data, has the potential to streamline haptic hardware selection for designers. This would bridge the knowledge gap and empower creators to develop richer and more effective touch experiences, building upon the foundation laid by existing repositories like Haptipedia \cite{seifi2019haptipedia} and the Haptiverse \cite{samithamby2021towards}.

In this paper, we propose a new system to address the limitations of current haptic device repositories and enhance the efficiency of haptic hardware recommendation systems. Leveraging the latest advancements in LLMs, we have developed an LLM-based haptic agent with a specific focus on Grounded Force Feedback (GFF) devices recommendation. GFF represents the most mature and extensively researched subset of haptic technology, offering a wide variety of devices across both research and commercial applications. This allows us to tailor the system to the specific needs of GFF devices, addressing the aforementioned challenges. The haptic agent consists of two primary subsystems: the creator-centric system and the end-user system  (example shown in Fig.\ref{fig:high_level}). The creator-centric system automates and facilitates the creation of new database entries by enabling hapticians to provide information about existing GFF devices sourced from research papers or product specifications, with minimal human verification. The end-user System is designed for designers seeking suitable GFF devices for their projects, allowing them to specify their requirements using natural language prompts. To handle the highly structured nature of GFF device specifications and the semantic richness of user queries, we propose a dynamic and adaptive Retrieval-Augmented Generation (RAG) approach. This method retrieves and reranks devices from both conditional and semantic searches to provide context in LLM prompts effectively.

In summary, this paper makes the following contributions:

\begin{itemize}
    \item We introduce an innovative haptic agent system leveraging LLMs specifically for GFF device recommendation. 
    % addressing the unique challenges and requirements of this mature subset of haptic technology.
    \item Utilizing the creator-centric system, we automate the process of populating the haptic device database, expanding it to include 114 entries.
    \item We propose a dynamic and adaptive RAG method tailored for tasks involving highly structured data and semantically rich user queries.
    \item We develop an online system that facilitates efficient searching functionalities and device discovery, empowering designers to find suitable GFF devices based on their specific requirements.
\end{itemize}

% \section{Related Work (May not needed since we are 6 page short paper)}

\section{Construction of A Haptic Agent}

% Our iterative process for creating Haptipedia comprised the two major sub system, creator system which convert the haptic device from source HTML and PDF into the haptic device databases, while the user system is mainly to contruct the LLM agent for the user queries for the end-point usage.

% This section details the iterative process for constructing a LLM-based haptic device recommendation agent. The process consists of two main sub-systems: \textit{Creator System} automates the conversion of haptic device information from raw source materials (HTML and PDF) into a structured database, eliminating the need for tedious manual data entry; \textit{User System} leverages a Large Language Model (LLM) to interpret user queries and recommend relevant haptic devices based on the constructed knowledge base.

This section outlines the iterative process for constructing an LLM-based haptic device recommendation agent. The process consists of two primary subsystems: the \textit{Creator-Centric System}, which automates the conversion of haptic device information from raw source materials into a structured haptic database, thereby eliminating the need for laborious manual data entry, and the \textit{End-User System}, which leverages a LLM to interpret user queries and recommend relevant haptic devices based on the constructed knowledge base.

\subsection{Creator-Centric System}
\label{sec:creator}
The creator-centric system is responsible for building the haptic device taxonomy database from raw source materials. The complete pipeline, as depicted in Fig. \ref{fig:framework}, comprises three main steps: 1) sourcing reliable haptic device material and converting it into processable information blocks; 2) constructing the haptic device taxonomy database using an LLM based on multi-source materials; 3) human validation of the haptic device taxonomy database to finalize it for the end-user system. These steps are further described below. By automating the data processing pipeline and leveraging the LLM for taxonomy construction, the creator-centric system significantly accelerates the development of the haptic device taxonomy database utilized in the end-user system.

% The creator system of the haptic agent is mainly to build the haptic device taxonomy database from raw source materials. The whole pipeline is shown in the \ref{fig:framework}, which mainly composed by 3 steps: a) sourcing the reliable haptic device material and convert it into processable info blocks; b) construct the haptic device taxonomy database by LLM based on the multi-sourcing materials; c) human validate the haptic device taxonomy database and finalize it for the user mode.
% By automatic the data processing pipeline and leveraging the LLM taxonomy construction, the creator system could significantly speed up the whole haptic device taxonomy database used for the user mode.

% \textbf{a) Haptic Device Sourcing.} 
\subsubsection{Haptic Device Sourcing} The creator-centric system generates a comprehensive haptic device taxonomy database from various sources, including academic literature and industry devices manuals and technical reports. Academic literature is obtained through a systematic search of relevant publications in major haptic journals and conferences, with primary input sources including original research papers (PDF) and existing online hyper text link (HTML). Industry resources are collected from prominent industry mechanisms through their official websites (HTML) and potential technical documentation (PDF). The system first parses these source files into distinct elements such as text blocks, tables, and images. Text and table blocks contain crucial device specification information necessary for extracting the taxonomy, which defines the categorization scheme for haptic device features. While image data is used as supplementary assets for taxonomy construction, capturing textual descriptions of visual information can complement the missing taxonomy of haptic devices, such as workspace.

\subsubsection{Taxonomy Construction} 
The taxonomy of the haptic device is primarily based on the definitions in \cite{seifi2019haptipedia}, which includes 41 machine attributes, 18 usage attributes, and 12 context attributes. To construct taxonomy for each haptic device, we explicitly adds supporting tags from the text, table and image blocks. To do that, we formulated prompts to extract a list of haptic device taxonomy topics mentioned in each section of the document (e.g., if that section refers to workspace or portability), as exemplified in listing Fig.\ref{fig:framework}b), and task the LLM model to answering them based on the extracted data. The prompt and model vary based on the input source: the LLaMA3 model is used to extract taxonomy from text and table strings, while the LLaVA-NeXT model extracts taxonomy from images. Each data block is requested to iterate through all the taxonomies required in the database. Subsequently, weighted voting is applied to aggregate the assigned tags, resulting in a consensus taxonomy for each device. The final aggregated taxonomy, along with the source blocks and relative metadata (such as source link), is used to initialize the SQL database.

Additionally, for research papers, the creator-centric system retrieves metadata using the Google Scholar API, which includes essential information such as title, authors, abstract, publication year, citation count, citation sources, and citations. This information serves as the foundation for building the knowledge base of academic haptic devices. For industry devices, an abstract summary of the devices is generated based on all text blocks.

\subsubsection{Human Review and Database Population} Following the automated stages, a human expert reviews the system-generated taxonomy to ensure its accuracy and alignment with the original source information. This human-in-the-loop approach guarantees the quality and reliability of the constructed knowledge base.

This step takes the taxonomy data from existing SQL database corresponding with source blocks. If the taxonomy data is corrected, the data would sync back with the SQL database, otherwise the annotator needs to correct the data based on the source blocks or the raw data source. As the result, the SQL database stores fundamental haptic device information like production year, degrees of freedom (DOF), and portability. Limited to the sourced documentation, the final SQL database is sparse, especially for the devices published with research papers. After this, the semantic taxonomy info including the title, abstract or portability are first transformed to text string and then encoded to high-dimensional vector representation of the complete device taxonomy information. The embedding would be stored at the vector database for the user semantic search.

\begin{figure}
\centering
\includegraphics[width=.43\textwidth, trim={20 23 20 23},clip]{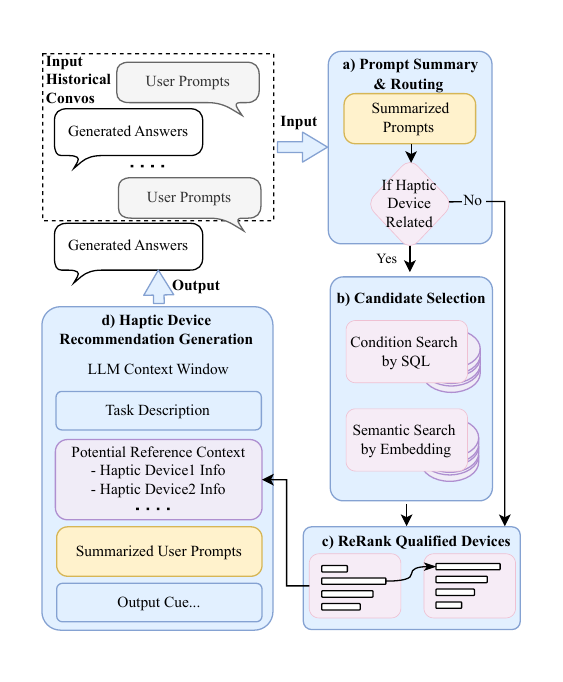}
   \caption{The pipeline of end-user system of haptic agent to interpret user queries and recommend relevant haptic devices based on the constructed knowledge base. Steps a)-d) represent the stages within the pipeline, with detailed descriptions available in Section \ref{sec:user}1-4) respectively.}
\label{fig:user_mode}
% \vspace{-0.3cm}
\end{figure}

\subsection{End-User System}
\label{sec:user}

The end-user system of haptic agent aims to establish a reliable haptic device recommendation agent by interpreting user queries. The entire process, illustrated in Fig.\ref{fig:user_mode}, comprises four primary steps: Initially (\textit{Step 1}), the system aggregates the end-user prompt and historical conversations into a summarized prompt and validates its relevance to haptic device recommendations. If the prompt is pertinent, the agent selects relevant haptic devices using both conditional and semantic searches within the haptic device database (\textit{Step 2}); if not, the system skips Step 2 and only references historically recommended devices. Subsequently (\textit{Step 3}), the candidates are reranked, and final device recommendations are made, incorporating necessary information into the prompts. Finally (\textit{Step 4}), the agent consolidates the selected haptic device information and the summarized user prompt to make the recommendation or response. These steps are further described below.

\subsubsection{Summarizing Prompts and Routing} 
The end-user system operates fundamentally on the basis of input prompts. To better understand user intent with contextual information, the system integrates the user prompt with historical conversations and interprets them to formulate a new, comprehensive query. This newly generated query undergoes an assessment to determine its relevance to the recommendation of a GFF haptic device, similar to the tagging system in creator-centric system shown in Fig.\ref{fig:framework}b).

Depending on the assessment outcome, the haptic agent adopts different approaches to candidate selection. 
If the query is unrelated to haptic device recommendations, the system aggregates historically recommended devices as candidates. 
% This aggregation ensures that users are provided with consistent and contextually relevant options based on their previous interactions, thereby enhancing the continuity and relevance of recommendations. 
Conversely, if the query is determined to be related, it is routed to the candidate selection process to identify potential new matches that align with the user's inquiry.
This routing process empowers the haptic agent to handle diverse queries effectively, ensuring a dynamic response to evolving user needs and facilitating accurate and pertinent recommendations throughout the conversation.

\subsubsection{Candidate Selection} To ensure precise haptic device recommendations, the system employs both conditional and semantic searches to identify potential new matches that align with the user's requirements and avoiding irrelevant or hallucinatory recommendations.

% In conditional searches, the agent generates multiple targeted queries to extract relevant haptic devices from the SQL database. These queries are based on the summarized user question and the underlying schema of the database. This SQL-based filtering provides detailed information necessary for further aggregation, focusing on devices with the core functionalities specified in the user's query, thereby ensuring the reliability and relevance of the recommendations. However, not all haptic device functionalities are captured within the SQL database. For these cases, the agent employs semantic similarity searches. This method involves projecting summarized user prompts into a high-dimensional embedding space and performing similarity searches within this space. By doing so, the system can identify devices stored in the vector database that align closely with the user's intent, ensuring that the recommended devices match the functional requirements specified by the user.

In conditional searches, the agent generates multiple targeted queries to extract relevant haptic devices from the SQL database. These queries are based on the summarized user question and the underlying schema of the database. This SQL-based filtering provides detailed information necessary for further aggregation, focusing on devices with the core functionalities specified in the user's query. However, not all haptic device functionalities are captured within the SQL database. For these cases, the agent employs semantic similarity searches. This method involves projecting summarized user prompts into a high-dimensional embedding space and performing similarity searches within this space. By doing so, the system can identify devices stored in the vector database that closely align with the user's intent, ensuring that the recommended devices match the functional requirements specified by the user.

% The combination of conditional and semantic searches 
This dual search approach allows the system to compile a comprehensive list of candidate devices. Conditional searches provide a structured and precise initial filter, ensuring that only devices meeting the basic technical specifications are considered. Semantic searches captures the nuanced functional attributes of haptic devices, ensuring alignment with the user's specific needs and preferences. %By leveraging both structured data and contextual understanding, it enhances the accuracy and relevance of the final recommendations.

\subsubsection{Reranking the Qualified Haptic Devices} 
Once the initial set of candidate devices is retrieved from the databases, candidate devices are reranked based on a combination of factors from both the SQL and vector databases. The ranking score for each haptic device, given the user prompt, is calculated using the following formula:

% the agent reranks the candidate devices based on a combination of factors extracted from both the SQL and vector databases. The ranking score of each haptic device given the user prompt is calculated based on the following equation,

\begin{equation}
RankScore_{Device} = \frac{N_{pos}}{N_{all}} + \cos (V_{Device}, V_{User})
\end{equation}
where \(N_{all}\) represents the number of technical specifications required by the user, and \(N_{pos}\) indicates how many of these specifications the device meets. \(\frac{N_{pos}}{N_{all}}\) shows the proportion of user-required attributes met by the device, while \(\cos (V_{Device}, V_{User})\) is the cosine similarity between the device's functionality embedding \(V_{Device}\) and the summarized user prompt \(V_{User}\).

% where $N_{all}$ represents the number of user required device's technical specifications, and $N_{pos}$ represents the number of haptic device's technical specifications meets the user needs. $N_{pos} / N_{all}$ represents by given haptic devices, what is the ratio of its attributes meets the user requires in the technical specifications. While $\cos (V_{Device}, V_{User})$ is the cosine similarity between the embedding of the device funcionality $V_{Device}$ and summerized user prompt $V_{User}$. 

Typically, the system shortlists the top N devices (N = 5) for final recommendation. This selection process guarantees that even if some requirements are only partially met, potential candidate devices are always available. This approach enhances the user experience by offering a diverse range of options that closely match the user's preferences and needs.

\subsubsection{Haptic Agent Response Generation} The final step involves the LLM aggregating all relevant information to generate a comprehensive response to the user's prompt. To achieve this, the system injects contextual information from the database into the LLM prompt, effectively combining the original query with detailed information about the shortlisted devices. This approach ensures that the responses are specifically tailored to the user's needs and preferences.

Specifically, the LLM's final prompt comprises four key elements: the task description, haptic device references, a summarized user prompt, and output cues. The task description defines the model's objectives, while output cues guide its response generation and format based on the provided inputs and context. To better align outputs with user needs, the system employs a few-shot learning approach to adapt task descriptions and output cues from multiple predefined options based on the summarized user prompt. The haptic device references organizes all pertinent data about the shortlisted devices after candidate selection and reranking, including their technical specifications and functional attributes. The summarized user prompt consolidates historical context and the current user inquiry from the \textit{Step 1}.

Following the generation of responses based on input prompts, post-processing techniques are employed to enhance output quality and record recommended devices. Since the output of the LLM responses only consists of summaries of recommended devices, the post process are adopted to allow users to explore detailed information about the recommended devices. For device recommendation-related questions, the final recommendations include haptic device IDs and links to source documentation or papers to ensure transparency and user-friendliness. Providing direct access to source information also enhances the credibility and reliability of the recommendations, as users can verify the details and assess the suitability of the devices for their specific needs. Concurrently, recommended devices are logged to refine suggestions progressively through ongoing user interactions, ensuring increasingly personalized and contextually relevant advice as conversations evolve.

\section{Experiments And Analysis}

% We conducted a study to accomplish two goals: evaluate if Haptic agent was a beneficial resource for novice and experienced hapticians, and identify further improvements to the system based on the ways novices and experts worked with and thought about authoring tools.

% The experiment are designed as following. We first let the hapticians try around 5 questions from the template, the template question are predefined by the creators and can give the user a sense of how the tooling is using. After that, we ask to the user to search for one haptic device they want to use for their potential project, and one of the haptic device they are interested in.

% After the experiment, we are ask online questionnaire was administered to both novices and experts. The participants completed the User Experience Questionnaire (UEQ) and indicated what one change they would most like to make to the classification system and to the website’s interface. The UEQ allows for a rating on six scales—Attractiveness, Perspicuity, Efficiency, Dependability, Stimulation, and Novelty—ranging from −3 to +3 where a value above 0.8 represents a positive evaluation. Then we introduce our LLM system shortly to the participants in 5 min and then take a 15 mins interview about what they thought about haptic device recommendation tool, described their prior experience for searching the haptic device, and what they think to improve the existing haptic device recommendation tool.

\subsection{Experiment Setup}
To evaluate the effectiveness of the proposed haptic agent and identify potential areas for improvement, we conducted a user study involving both novice and experienced hapticians. The study aimed to determine if the haptic agent serves as a valuable resource for users with varying levels of expertise, assessing its usability and benefits in helping them find suitable haptic devices. Additionally, we sought to gather user feedback to gain insights into how different user groups interact with and perceive the system. This feedback is essential for identifying areas for improvement the system to better meet the diverse needs of its users.

% To evaluate the effectiveness of the proposed haptic agent and identify potential areas for improvement, we conducted a user study with novice and experienced hapticians. The study aimed to achieve two primary goals: \textbf{Assess the Usability and Benefit of Haptic Agent}: We sought to determine if Haptic Agent serves as a valuable resource for both novice and experienced users. This involved evaluating how effectively the system assisted users in finding suitable haptic devices for their needs.
% \textbf{Gather User Feedback for System Improvement}: We aimed to gain insights into how novices and experts interact with and perceive the system. This feedback would be crucial for identifying areas for improvement and tailoring the system to better meet the needs of diverse user groups.

The study recruited a balanced cohort of 13 participants, each with varying levels of experience with haptic devices, as illustrated in Fig.\ref{fig:haptic_exp}a). Among them, two participants were the creators of the system under study. Fig.\ref{fig:haptic_exp}b) presents the distribution of participant expertise. The group included seven novice hapticians, defined as individuals with less than two years of experience in working with haptic devices. Though they were not experts in the selection of specific devices, these participants possessed a basic understanding of GFF devices. The remaining six participants were experienced hapticians, each with a robust background in haptics and a proven track record in utilizing various haptic devices within research or professional settings; two of these experienced participants are currently employed in the industry.

% The study recruited a balanced group of participants of \textbf{9} people with varying levels of haptic device experience shown in the Fig. \ref{fig:haptic_exp} a), where two of them are creator of the system. Fig. \ref{fig:haptic_exp} b) shown the distribution of the participants. There are \textbf{5} novice haptician consisted of individuals with less than 2 years experience working with haptic devices. Ideally, participants would have a basic understanding of haptics technology but would not be experts in selecting specific devices. The other 4 participants are comprised experienced hapticians with a strong background in haptics and a proven track record of working with various haptic devices in research or professional settings, where 2 of them are from the industry.
% The target sample size should be statistically significant for the chosen analysis methods. Common practices in HCI research suggest a minimum sample size of 15-30 participants per user group, depending on the complexity of the system and the desired effect size. Recruitment strategies could include online forums dedicated to haptics research, university research communities, or industry conferences.

\subsection{Experimental Procedure}

The user study was conducted in a structured format to evaluate interactions with the haptic agent and to collect participant feedback. Participants first received an overview of the study objectives and a brief introduction to the haptic agent, including five pre-defined questions to illustrate its capabilities. They then completed two search tasks using the haptic agent: \textit{Task 1} involved locating a suitable commercial haptic device for a potential project, while \textit{Task 2} required finding a haptic device from research papers that sparked their interest or appeared relevant for further investigation. These tasks aimed to assess the system's ability to recommend suitable devices based on user needs. To facilitate direct comparison, participants were also asked to perform these tasks using existing resources such as Google search, Hapticpedia \cite{seifi2019haptipedia} or GPT4-Turbo. Throughout the experiments, the haptic agent adopted GPT-3.5 Turbo as the LLM backend to enhance efficiency and simplify deployment.

% First, they searched for a \textit{commercial haptic device} suitable for a project, testing the system's recommendation accuracy. Second, they searched for a \textit{haptic device from research papers} of interest or relevance, assessing the system's ability to discover innovative devices. 

After completing the tasks, participants filled out a standardized User Experience Questionnaire (UEQ) and optionally participated in a 15-minute semi-structured interview to provide detailed qualitative feedback on their experiences with the haptic agent.

\subsection {Experimental Results and Analysis}

The UEQ used a multi-scale Likert format, ranging from -3 (strongly disagree) to +3 (strongly agree), to assess participants' subjective experiences with the system \cite{sullivan2013analyzing}. The UEQ's subscales included Attractiveness, Perspicuity, Efficiency, Dependability, Stimulation, and Novelty, providing insights into various aspects of user perception such as aesthetics, ease of use, effectiveness, reliability, engagement, and innovativeness. For each aspect in UEQ, two questions were formulated: one to capture users' feelings about the haptic agent system itself and another to facilitate comparisons with existing systems for a comprehensive analysis.

% The UEQ employed a multi-scale Likert format, ranging from -3 (strongly disagree) to +3 (strongly agree), to assess participants' subjective experiences with the system \cite{sullivan2013analyzing}. The UEQ's subscales—Attractiveness, Perspicuity, Efficiency, Dependability, Stimulation, and Novelty—were designed to provide insights into various aspects of user perception, including the system's aesthetics, ease of use, effectiveness, reliability, engaging nature, and innovativeness. For each of these aspects, we formulated two distinct questions: one to capture the users' feelings about the Haptic Agent system itself, and another to facilitate comparisons with existing systems, thereby enabling a more comprehensive analysis.

\subsubsection{Benchmark with Established UEQ}
\begin{figure}
\centering
% \includegraphics[width=.35\textwidth, trim={10 10 10 10},clip]{figs/haptic_high_framework.pdf}
% \vspace{-.1cm}
 \includegraphics[width=.40\textwidth, trim={20 37 25 37},clip]{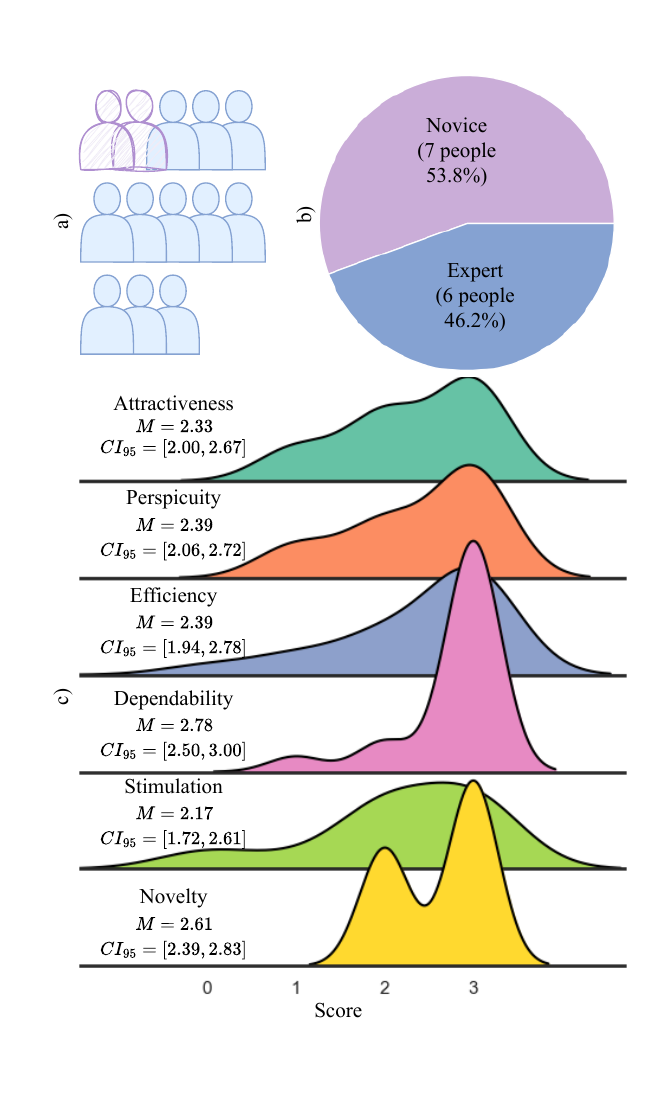}
   \caption{Overview of Experimental Setups and User Experience Analysis. a) Composition of experiment participants, with creators(2) marked in purple and users(11) in blue; b) Distribution of participants by years of experience with haptic devices; c) Density distribution of UEQ across six subscales, with responses ranging from -3 (strongly disagree) to +3 (strongly agree) - responses below 0 are not shown as there were no answers in this range.}
\label{fig:haptic_exp}
% \vspace{-.4cm}
\end{figure}

The results of the UEQ for the haptic agent are presented in Fig.\ref{fig:haptic_exp}c) with corresponding mean and 95\% confidence interval marked as $M$ and $CI_{95}$, respectively.
These intervals were calculated using bootstrap sampling of the UEQ rating data with 10K iterations for each aspect, rather than fitting them into a specific distribution due to diverse distributions across aspects. The lower bound of $CI_{95}$ is higher than 2 for most fields, except for efficiency and stimulation. Efficiency had the largest variance (0.83) among the six sub-scales, primarily related to participants' familiarity with LLM. Stimulation has the lowest confidence interval among all fields, where users suggested that incorporating visual elements into the haptic agent could improve interaction and satisfaction.
Comparing these results with benchmarks from \cite{schrepp2017construction}, which assess established product evaluations based on mean UEQ scales, the proposed haptic agent ranks in the top 10\% across all categories: Attractiveness ($>$1.75), Perspicuity ($>$1.78), Efficiency ($>$1.9), Dependability ($>$1.65), Stimulation ($>$1.55), and Novelty ($>$1.4). This performance highlights the system's superior user experience in key dimensions of usability and satisfaction.

\subsubsection{Benchmark with Existing Haptic Device Searching Tool} The performance of the haptic agent surpasses existing systems. Specifically, when comparing self-evaluation questions with those comparing the haptic agent to existing systems, the mean scores of the comparing group are higher across nearly all subscales, except for Stimulation. The differences in means are as follows: Attractiveness ($M^{diff}_a = 0.67$), Perspicuity ($M^{diff}_{p} = 0.11$), Efficiency ($M^{diff}_{e} = 0.34$), Dependability ($M^{diff}_{d} = 0.22$), Stimulation ($M^{diff}_{s} = -0.34$), and Novelty ($M^{diff}_{n} = 0.11$). These results indicate that participants rated the haptic agent higher when making comparisons. This suggests that despite potential optimization issues, the haptic agent still outperforms the existing system.
Though, using the Mann-Whitney U test to compare the two question groups, a significant difference was found only in the attractiveness subscale (p=0.047). No significant differences were found for Perspicuity (p=0.882), Efficiency (p=0.577), Dependability (p=0.540), Stimulation (p=0.774), and Novelty (p=0.676).

To mitigate bias from different information sources or haptic backgrounds, the Mann-Whitney U test was conducted between creators and users, revealing no significant differences. The p-values for each subdomain are as follows: Attractiveness (p=0.09), Perspicuity (p=0.44), Efficiency (p=0.15), Dependability (p=0.43), Stimulation (p=0.06), and Novelty (p=0.15).
Similarly, the Mann-Whitney U test was applied to compare novice and expert groups, with no significant differences observed. The p-values are: Attractiveness (p=0.80), Perspicuity (p=0.62), Efficiency (p=0.20), Dependability (p=0.84), Stimulation (p=0.47), and Novelty (p=0.32). These results demonstrate that the haptic agent performs well across different groups.

% and between novice and expert groups. 

% only the attractiveness field showed a significant difference (p=0.047), indicating no significant difference in attractiveness compared to existing systems despite high self-evaluation ratings. No significant differences were found for perspicuity (p=0.882), efficiency (p=0.577), dependability (p=0.540), stimulation (p=0.774), and novelty (p=0.676). To avoid bias from different information sources or haptic backgrounds, the Mann-Whitney U test was conducted between creators and users, and between novice and expert groups, revealing no significant differences between these groups as well.

% More specifically, by comparing with the rating between the self-evaluate questions and the comparing with existing system questions, only the attractiveness field has significant difference (p=0.047), which shows there is no significant attractiveness comparing with existing systems even though the self-evaluate rating is high. No significant difference was present for perspicuity (p=0.882), efficiency (p=0.577), dependability (p=0.540), stimulation (p=0.774) and novelty (p=0.676). Also, to avoid the result is bias between the different information source or the haptic background, we also adopt the Mann-Whitney U test between the creators and users group, and between the novice and expert group. And there is no significant difference was present between these two groups as well.

\subsubsection{User Insights}
Based on responses from open-ended questions in the questionnaire and interviews, users overwhelmingly found the haptic agent to be innovative and reliable, capable of delivering accurate information without errors. This was particularly evident in \textit{Task 2}, focused on locating haptic research papers, where the agent outperformed existing generative AI by providing more efficient and reliable results. Users also noted enhanced efficiency in finding haptic devices in both tasks, especially those less familiar with haptics or unsure of specific device criteria. Feedback highlighted the agent's ability to start with partially matched devices based on vague prompts and refine recommendations dynamically during interactions. However, concerns were raised about the purely text-based interface's reduced efficiency and engagement compared to image-supported platforms like Hapticpedia \cite{seifi2019haptipedia}. Additionally, the haptic agent's performance lagged behind Hapticpedia in interaction speed due to multiple LLM calls, where this issue is expected to improve with advancements in LLM algorithms, deployment optimizations, and hardware enhancements \cite{dong2024llm}. Users with limited LLM experience reported that they needed more time and effort to find effective prompting patterns, which potentially reduced efficiency, especially for initial queries.

\section{Conclusion}

In conclusion, this paper presents a new system to addressing the limitations of current haptic device repositories by leveraging advancements in LLMs to develop an intelligent haptic device recommendation agent focused on GFF devices. Our system automates the creation of a structured database from diverse sources and provides precise, contextually relevant recommendations through a dynamic RAG method. User evaluations indicate superior performance in usability and satisfaction compared to existing systems.
% Future work will focus on embedding additional types of haptic devices and expanding the database to further enhance the system's capabilities and utility.

% \bibliographystyle{ieee_fullname}
% \bibliography{egpaper}
\bibliographystyle{plain}
\bibliography{egpaper}

\renewenvironment{IEEEbiography}[1]
  {\IEEEbiographynophoto{#1}}
  {\endIEEEbiographynophoto}

% \vspace{-1.2cm}
\begin{IEEEbiography}{Yang Liu}
is a PhD Candidate at the University of Ottawa and a member of IEEE. His research interests include artificial intelligence and multimedia. Contact him at yliu344@uottawa.ca.
\end{IEEEbiography}

% \vspace{-1.2cm}
\begin{IEEEbiography}{Haiwei Dong}
is a Director and Principal Researcher at Huawei Canada, and an Adjunct Professor at the University of Ottawa. His research interests include artificial intelligence, multimedia, Metaverse, and robotics. He is a Senior Member of IEEE, a Senior Member of ACM, and a registered Professional Engineer in Ontario. Contact him at hdong@uottawa.ca.\end{IEEEbiography}

% \vspace{-1.2cm}
\begin{IEEEbiography}{Abdulmotaleb El Saddik} is a Distinguished Professor at the University of Ottawa. His research focus is on multimodal interactions with sensory information in smart cities. He is a Fellow of Royal Society of Canada, a Fellow of the Engineering Institute of Canada, a Fellow of the Canadian Academy of Engineers, a Fellow of IEEE and an ACM Distinguished Scientist. He received IEEE I\&M Technical Achievement Award and IEEE Canada Computer Medal. Contact him at elsaddik@uottawa.ca.
\end{IEEEbiography}
\end{document}